\documentclass[prx,twocolumn, superscriptaddress,nofootinbib,floatfix]{revtex4-1}
\usepackage{footmisc}
\usepackage{graphicx}
\usepackage{natbib,ulem,soul}
\usepackage{bbold}
\usepackage{amsmath,epsfig,color,amssymb}

\usepackage{xcolor}
\definecolor{dark-red}{rgb}{0.4,0.15,0.15}
\definecolor{dark-blue}{rgb}{0.15,0.15,0.4}
\definecolor{medium-blue}{rgb}{0,0,0.5}
\usepackage[bookmarks=false]{hyperref}
\usepackage[all]{hypcap}
\usepackage{cleveref}
\hypersetup{
    colorlinks, linkcolor={dark-red},
    citecolor={dark-blue}, urlcolor={medium-blue}
} 

\newcommand{\ssp}{\hspace{0.4pt}}

\newcommand{\ket}[1]{\lvert\, #1\, \rangle}

\newcommand{\hc}{\mathrm{h.c.}}

\newcommand{\nn}{n}
\newcommand{\nno}{N}
\newcommand{\cc}{c}

\graphicspath{{./Figures/}}

\begin{document}

\author{Armin Rahmani}
\affiliation{Department of Physics and Astronomy and Advanced Materials Science and Engineering Center, Western Washington University, Bellingham, Washington 98225, USA}
\affiliation{Kavli Institute for Theoretical Physics, University of California, Santa Barbara, California 93106, USA}

\author{Kevin J. Sung}
\affiliation{Google AI Quantum, Santa Barbara, CA, USA}
\affiliation{Department of Electrical Engineering and Computer Science,
University of Michigan, Ann Arbor, MI 48109, USA}

\author{Harald Putterman}
\affiliation{Google AI Quantum, Santa Barbara, CA, USA}

\author{Pedram Roushan}
\affiliation{Google AI Quantum, Santa Barbara, CA, USA}

\author{Pouyan Ghaemi}
\email{pghaemi@ccny.cuny.edu}
\affiliation{Physics Department, City College of the City University of New York, New York, NY 10031}
\affiliation{Graduate Center of the City University of New York, New York, NY 10016}

\author{Zhang Jiang}
\email{qzj@google.com}
\affiliation{Google AI Quantum, Santa Barbara, CA, USA}


\title{Creating and manipulating a Laughlin-type \texorpdfstring{$\nu=1/3$}{v=1/3} fractional quantum Hall state on a quantum computer with linear depth circuits}

\begin{abstract} 
Here we present an efficient quantum algorithm to generate an equivalent many-body state to Laughlin's $\nu=1/3$ fractional quantum Hall state on a digitized quantum computer. Our algorithm only uses quantum gates acting on neighboring qubits in a quasi one-dimensional setting, and its circuit depth is linear in the number of qubits, i.e., the number of Landau orbitals in the second quantized picture. We identify correlation functions that serve as signatures of the Laughlin state and discuss how to obtain them on a quantum computer. We also discuss a generalization of the algorithm for creating quasiparticles in the Laughlin state. This paves the way for several important studies, including quantum simulation of nonequilibrium dynamics and braiding of quasiparticles in quantum Hall states.
\end{abstract}

\maketitle

\section{Introduction} 

Understanding the properties of strongly interacting electrons is a long-standing challenge in condensed-matter physics~\cite{quintanilla_strong-correlations_2009,basov_electrodynamics_2011}. The computation time required for the numerical solution of strongly correlated quantum many-body models on classical computers grows exponentially with the system size. As proposed by Feynman in 1982, the use of a quantum computer to simulate a quantum system can circumvent this difficulty~\cite{georgescu_quantum_2014,wecker_solving_2015,reiher_elucidating_2017,babbush_low-depth_2018}. Recent advances in superconducting qubits~\cite{barends_superconducting_2014, sheldon_procedure_2016, Arute2019} and trapped-ion qubits~\cite{lanyon_universal_2011, blatt_quantum_2012}, among other platforms, have brought us close to this goal. Concurrently, progress has been made on algorithms to simulate strongly correlated quantum systems~\cite{jiang_quantum_2018,PhysRevX.8.041015,mcclean_theory_2016, childs_nearly_2019}. 

Existing and near-term quantum hardware provide an unprecedented opportunity for creating strongly correlated states of quantum matter that can be controlled and manipulated to a high degree of precision. The ability to create such states on general-purpose digitized quantum devices (rather than customized single-purpose systems that physically emulate the associated Hamiltonians) can foster rapid progress in quantum simulation of correlated quantum systems. Such advances, however, rely on the development of efficient quantum algorithms. 
Recently, different algorithms have been proposed for the use of quantum computers to study topological phases as well as the Hubbard model~\cite{kivlichan_improved_2019,PhysRevX.8.041015,wecker_solving_2015, jiang_quantum_2018}, the latter corresponding to a strongly interacting system that is believed to capture the essential properties of cuprate high-temperature superconductors. 

Another class of correlated electron states are fractional Hall phases~\cite{kaicher_roadmap_2020}. In the latter system, the kinetic energy of electrons is suppressed by an external magnetic field, and, as a result, even the screened interaction is strong in comparison with the kinetic energy. It has been experimentally shown that the correlated electronic states that emerge in these settings would harvest novel properties such as fractional quasiparticle charges. These experimental results have first been explained by Laughlin through the prediction of the form of a many-body wave function~\cite{PhysRevLett.50.1395}. Numerical methods such as exact diagonalization have further confirmed the form of the Laughlin wave function as the many-body wave function corresponding to fractional Hall phases. Diverse types of correlated states seem to develop at different filling fractions of the Landau level. The types of fractional Hall states predicted theoretically are more numerous than those that have been experimentally realized. Many of the theoretically predicted phases would display novel phenomena such as non-Abelian braiding. Despite extensive effort, many of these demanding theoretical predictions have not been experimentally realized. The generation of fractional Hall states on quantum computers would provide a highly controllable platform to realize novel phenomena that have been predicted theoretically, on an actual quantum wave function. Fractional Hall states can result from long-ranged electron-electron interactions but to generate them efficiently on quantum computers, we need unitary operators that act only on nearest-neighbor qubits. 

In this paper, we present a quantum algorithm to generate a $\nu=1/3$ fractional quantum Hall state in the same topological class as Laughlin's wave function on a digitized quantum computer.  Our algorithm works in the second-quantized representation of Landau orbitals in each Landau level, where the parent Hamiltonian is one dimensional (1D) and gapped (with a matrix-product ground state). 
Our proposal requires $N$ qubits, where $N$ is the number of Landau orbitals, and employs a quantum circuit of depth $N/3 + 3$. It opens up new avenues to simulating novel quantum phenomena in fractional quantum Hall systems on a digitized quantum computer. 

In Sec.~\ref{sec:spin_chain}, we discuss the 1D parent Hamiltonian of the Laughlin state in the second quantization picture.  In Sec.~\ref{sec:quantum_circuit}, we first introduce a reduced representation of the Laughlin state using only one third of the qubits. This state represents a fractional quantum Hall state with $\nu=1/3$ and is equivalent to the Laughlin's wave function, which itself is not the exact ground state for a realistic Coulomb Hamiltonian. The effect of the truncation is benign, given the fast Gaussian decay of the pseudopotentials. We then prepare the Laughlin state in the reduced representation before converting it back to the original representation.  In Sec.~\ref{sec:verification}, we discuss how to verify the generated Laughlin state by evaluating certain correlation functions.  In Section~\ref{sec:quasi_particle}, we discuss how to create quasiparticles in the Laughlin state. We present our conclusion in Sec.~\ref{sec:conclusion}.

\section{Fermion-chain models for fractional Hall effect}
\label{sec:spin_chain}
When a perpendicular magnetic field is applied to a two-dimensional (2D) electron gas, the kinetic energy of electrons is suppressed and they form a set of discrete Landau levels. The degeneracy of each level is determined by the number of magnetic flux quanta passing through the system, while the gap between the Landau levels is directly controlled by the size of the magnetic field. Within each Landau level, the electron-electron interaction is the sole energy scale and it determines the many-body ground state. Since the degenerate set of states could be labeled with a single quantum number (e.g., angular momentum), an effective 1D model could be applied to study the quantum states within each Landau level. The main challenge, then, is that the electrons in all states in the 1D chain interact with each other. To represent the interaction potentials between the states in one Landau level, we need to choose a basis with which to represent the degenerate states within the lowest Landau level. A commonly used method is to consider states for electrons on a torus. The general form of the Hamiltonian will then correspond to ~\cite{Seidel2005,Flavin2011,nakamura_exactly_2012}
\begin{equation}\label{eq:hamil}
    \mathcal{H}=\sum_{i=0}^{N-1} \sum_{k>|m|}V_{km} c^\dagger_{i+m}c^\dagger_{i+k}c_{i+m+k}c_i\,
\end{equation}
where $c_i$ is the annihilation operator acting on orbital $i$ and $N=\frac{L_1L_2}{2\pi}$ identifies the number of available states in each Landau level. With periodic boundary conditions, we identify $c_{N+j}$ with $c_j$. $L_1$ and $L_2$ are the circumferences of the
torus. The interaction potential $V_{km}$ corresponds to the projected Coulomb interaction in the orbitals within each Landau level. Generally speaking, $V_{km}$ is nonzero for all pairs of orbitals but considering specific limits, such as the thin-torus limit, the interaction potential can be simplified. For example, the form of the interaction potential that leads to the Laughlin wave function,
\begin{equation}
V_{km}\propto (k^2-m^2) e^{-2\pi^2(m^2+k^2)/L_1}
\end{equation}
gives local interaction in the limit of $L_1\rightarrow 0$ and has a density-wave ground state $|100100\dots\rangle$, where each occupied site is followed by two unoccupied sites. 

Following~\cite{nakamura_exactly_2012}, we truncate the above Hamiltonian to a minimal model with a fractional-quantum-Hall-type ground state:
\begin{align}
    H &=\sum_{j=0}^{\nno-1} V_{10} \nn_{j+1} \nn_{j+2} +V_{20} \nn_j\nn_{j+2} +V_{30}\nn_j\nn_{j+3}\nonumber\\
    &\hspace{1cm} + \sqrt{V_{10}V_{30}}\big(\cc_j^\dagger\ssp\cc_{j+3}^\dagger\cc_{j+2}\ssp\cc_{j+1} +\hc\big)\,,\label{eq:1d_fqh_a}
\end{align}
where $\nn_j = \cc_j^\dagger\cc_j$. 

Applying transformations that correspond to enlarging $L_1$ transforms the density-wave ground state into an entangled state corresponding to a Laughlin-type state. Even in the thick-torus limit, the state has an important simplifying feature: it can be represented by a matrix product state that captures all the salient properties of the Laughlin state~\cite{Seidel2005,Flavin2011,nakamura_exactly_2012}.  





\section{Quantum circuit} 
\label{sec:quantum_circuit}

For the Laughlin $\nu=1/3$ state, it has been shown in Ref.~\cite{nakamura_exactly_2012} that the (unnormalized) ground state of Hamiltonian~\eqref{eq:hamil} can be obtained by the action of a nonunitary operator on a direct-product state as
\begin{equation}\label{eq:nonU}
    |\psi\rangle={\cal N}\prod_{j=0}^{N-3}(1-t
c^\dagger_{j+1}c^\dagger_{j+2}c_{j+3}c_j)|100100100\ldots\rangle,
\end{equation}
where the parameter $t$ is given by $t=\sqrt{V_{30}/V_{10}}$. The factor $\cal N$ is an overall normalization constant.

It is convenient to map the fermions to qubits through a Jordan-Wigner transformation, which maps the fermion occupation number $n_j=c^\dagger_jc_j$ to spin operators $\sigma^z_j=1-2n_j$. The two levels of each qubit satisfy $\sigma^z_j|0\rangle_j=|0\rangle_j$, $\sigma^z_j|1\rangle_j=-|1\rangle_j$, and the fermion creation and annihilation operators are given by
\begin{equation}
    c_j=\left(\prod_{k<j}\sigma^z_k\right)\sigma^-_j,\quad c^\dagger_j=\left(\prod_{k<j}\sigma^z_k\right)\sigma^+_j,
\end{equation}
where $\sigma^+_j|0\rangle_j=|1\rangle_j$, $\sigma^-_j|1\rangle_j=|0\rangle_j$, $\sigma^+_j|1\rangle_j=0$, and $\sigma^-_j|0\rangle_j=0$.

Using the anticommutation of fermion operators, we can then write the squeezing operator $c^\dagger_{j+1}c^\dagger_{j+2}c_{j+3}c_j=\sigma^+_{j+2}\sigma^z_{j+2}\sigma^-_{j+3}\sigma^+_{j+1}\sigma^z_{j}\sigma^-_{j}$ that corresponds to moving the electrons on the effective 1D chain toward each other \cite{PhysRevLett.100.246802,PhysRevB.95.245123}. The state can then be written as $|\psi\rangle={\cal N}\prod_j(1-t
S_j)|100100100\ldots\rangle$ 
\begin{equation}
S_j=\sigma^+_{j+1}\sigma^+_{j+2}\sigma^-_{j+3}\sigma^-_j,
\end{equation}
where we use $\sigma^+\sigma^z=\sigma^+$ and $\sigma^z\sigma^-=\sigma^-$.


To construct a unitary operator that creates the same state, we note that blocks of three consecutive qubits effectively serve as a reduced qubit, indicating whether or not a block is squeezed. The initial state has no squeezed blocks and can be represented by a sequence of zeros in the reduced space as $|100100100100\rangle\to|\mathbb{0000}\rangle$. As another example, a state with two squeezed blocks $|\mathbb{1010}\rangle$ represents $|011000011000\rangle$. 
The squeezing operator associated with a block acts on the three qubits in the block and the first qubit of the next block. A squeezing operator transforms the qubits in a three-qubit block as well as the first site of the next block according to $|1{ 0}0,1\rangle\to|0{ 1}1,0\rangle$.  Therefore, when the squeezing operator acts on a given block, its first qubit as well as the first qubit of the next three-qubit block will be set to $0$. We note that $S_j$ annihilates the state unless the four qubits it acts upon are in the $1001$ configuration. Thus, if a block is squeezed, the application of the squeezing operator on the neighboring block to its left (right) will annihilate the state. Thus, a squeezed block cannot have any neighboring blocks that are also squeezed. For example, the reduced representation on two neighboring blocks are
\begin{eqnarray}
    |\mathbb{1}_k\rangle &=& |0_{3k}1_{3k+1}1_{3k+2}\rangle\\
    |\mathbb{1}_k\mathbb{0}_{k+1}\rangle &=& |0_{3k}1_{3k+1}1_{3k+2}, 0_{3k+3}0_{3k+4}0_{3k+5}\rangle\\
        |\mathbb{0}_k\mathbb{0}_{k+1}\rangle &=& |1_{3k}0_{3k+1}0_{3k+2}, 1_{3k+3}0_{3k+4}0_{3k+5}\rangle
\end{eqnarray}

The number of states in the superposition~\eqref{eq:nonU} is then given by the Fibonacci number, which grows exponentially with the system size\footnote{We can see this by noting the recursive nature of the number of states, $F(m)$  for $m$ blocks. If the first register is $\mathbb{1}$, the second register must be $\mathbb{0}$ and we have $F(m-2)$ possibilities for the remaining registers. If the first register is $\mathbb{0}$ on the other hand, there is no restriction on the other resisters and we have $F(m-1)$ possibilities. Therefore, $F(m)=F(m-1)+F(m-2)$.}. The amplitude of a state in the unnormalized superposition is then $(-t)^P$, where $P$ is the number of $\mathbb 1$s in the reduced space of registers associated with the blocks. 

The nonunitary operator $(1-tS_{3k})$ acting on $\ket{\mathbb{0}_k}$ creates $\ket{\mathbb{0}_k}-t\ket{\mathbb{1}_k}$ when the ($k+1$)-th register is $\mathbb 0$ and it leaves the $|\mathbb{0}_k\rangle$ unchanged when the ($k+1$)-th register is $\mathbb 1$. Since a nonunitary operator cannot be implemented on a quantum device, we consider the unitary operator on the $k$-th register
\begin{equation}\label{eq:uk}
    U_k=e^{\phi_k(S_{3k}-S^\dagger_{3k}),}
\end{equation}
which similarly acts as identity when the ($k+1$)-th register is $\mathbb 1$. However, when the ($k+1$)-th register is $\mathbb 0$, it creates a superposition 
\begin{equation}
 U_k  |\mathbb{0}_k\mathbb{0}_{k+1}\rangle= \cos(\phi_k)|\mathbb{0}_k\mathbb{0}_{k+1}\rangle+\sin(\phi_k)|\mathbb{1}_k\mathbb{0}_{k+1}\rangle
\end{equation}
 
 Our approach to creating the state~\eqref{eq:nonU} uses the following sequence of unitary operators
\begin{align}
 \ket{\psi} = U_{N/3-1}(\phi_{N/3-1})\cdots U_{1}(\phi_{1})\, U_{0}(\phi_{0})\,\ket{\mathbb{0}\cdots \mathbb{00}}\,.
\end{align}
By choosing the appropriate angle $\phi_k$ obtained from the recursion relation 
\begin{equation}
  \phi_{k-1}=\arctan\left[-t\cos(\phi_k)\right],
\end{equation}
with boundary condition $\phi_{N/3-1}=\arctan(-t)$, we can ensure that the resulting normalized state is the same as Eq.~\eqref{eq:nonU}.

The recursion provides the correct factor of $(-t)^P$ for a state with $P$ squeezed registers $\ket{\mathbb 1}$. The amplitude of the $(k-1)$-th register being in state $\ket{\mathbb 0}$  is reduced by a factor of $\cos\phi_j$ if the state of the $k$-th register is  $\ket{\mathbb 0}$; otherwise, it remains the same. On the other hand, every register with state $\ket{\mathbb 1}$ obtains a factor of $\sin \phi_j$ during the squeezing operation. 
Therefore, the squeezing angles on two consecutive blocks must satisfy
\begin{align}
    \frac{\sin(\phi_{k-1})}{ \cos(\phi_k) \cos(\phi_{k-1})}=-t\,,
\end{align}
to create the superposition $\ket{\mathbb{0}_{k-1}}-t\ket{\mathbb{1}_{k-1}}$.  For example, we consider the initial state $|100100100\rangle=|\mathbb{000}\rangle$. The resulting squeezed state is $ U_2U_1U_0|\mathbb{000}\rangle=c_0c_1c_2|\mathbb{000}\rangle+c_0c_1s_2|\mathbb{001}\rangle+c_0s_1|\mathbb{010}\rangle
     +s_0c_2|\mathbb{100}\rangle+s_0s_2|\mathbb{101}\rangle$,
where we use the shorthand notation $c_j\equiv \cos(\phi_j)$ and $s_j\equiv \sin(\phi_j)$. The condition $\tan \phi_2=\frac{\tan \phi_1}{\cos \phi_2}= \frac{\tan\phi_0}{\cos\phi_1}=-t$ needs to be satisfied to create the desired superposition $\ket{\mathbb{0}}-t\ket{\mathbb 1}$ for all blocks.

\begin{figure}[htp]
	\includegraphics[width=8cm]{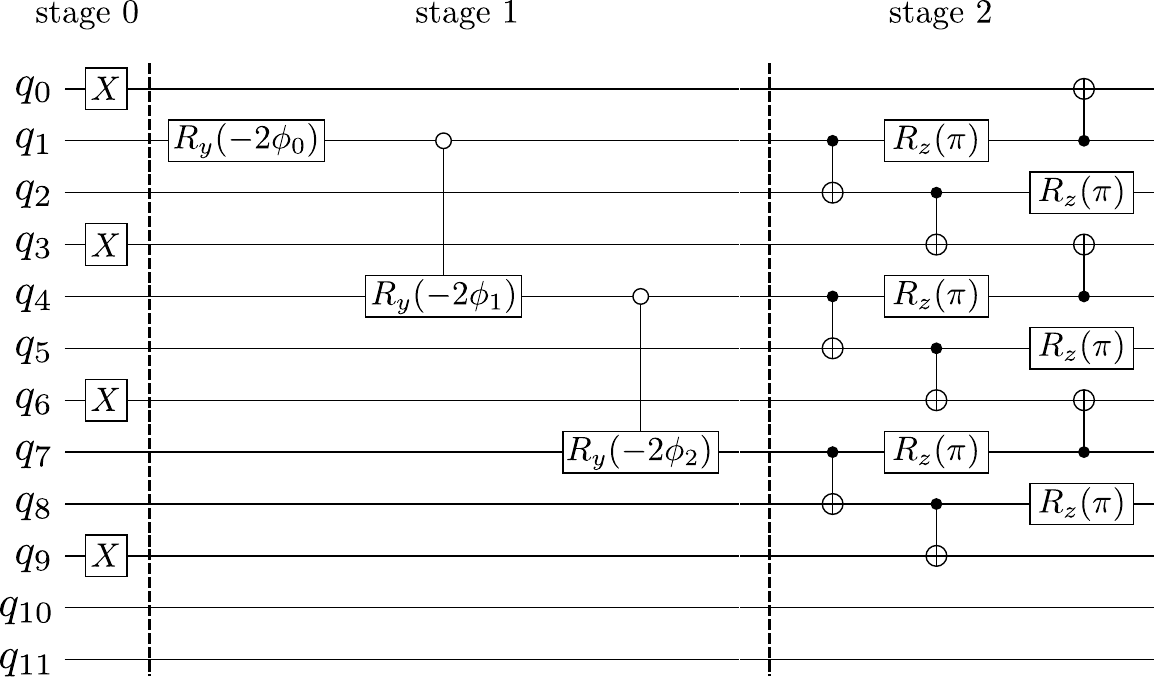}
	\caption{The full quantum circuit in three stages.}
	\label{fig:circ}
\end{figure}

We now present our algorithm in the form of a unitary quantum circuit acting on the vacuum state $|000\dots 0\rangle$ of $3m$ superconducting qubits. The circuit depth scales linearly with the number of qubits, with a small enough coefficient to allow the implementation of the algorithm on existing quantum hardware. The efficiency of the algorithm relies on two key observations. First, as discussed above, we cannot squeeze two neighboring three-qubit blocks.  The second observation is that the middle site of a block fully encodes whether or not a block is squeezed. It is $|0\rangle$ ($|1\rangle$) before (after) squeezing. As shown in Fig. (\ref{fig:circ}), our algorithm starts by creating the initial direct-product state $|100100100\dots\rangle$ using $X$ gates in stage 0.
Stage 1 is the core of the algorithm, which applies the squeezing operators on the middle site of each three-qubit block. Initially, none of the blocks are squeezed, so for the first block we simply apply a unitary rotation $R_y(\theta)\equiv e^{-i\theta\sigma^y/2}$ on the middle site of the first block. For the second block, however, we apply a controlled rotation. If the middle site of the first block is 1, a neighboring block is squeezed and we should not act on the second block. The middle site of block $k$ acts as a control register for a $R_y(-2\phi_{k+1})$ rotation applied to the middle site of block $k+1$. In this implementation, we make use of a controlled phase gate, \small{CNOT}, and simple single-qubit gates. Since the gates are applied sequentially, we only need to control each block by the previous block.

At the end of stage 1, the middle qubit of each block has the same amplitude as the corresponding reduced register. In the final stage, we use control gates to fix the states of the neighboring qubits according to whether or not a squeezing has occurred.

Therefore, in the first stage of the algorithm we just act on the middle site of each block to construct the overall structure of the superposition. In the following stage, we use control gates to fix the nearby qubits. Several simplifications, utilizing the special structure of the initial state, are used to eliminate redundant gates.

\begin{figure}[htp]
	\includegraphics[width=5cm]{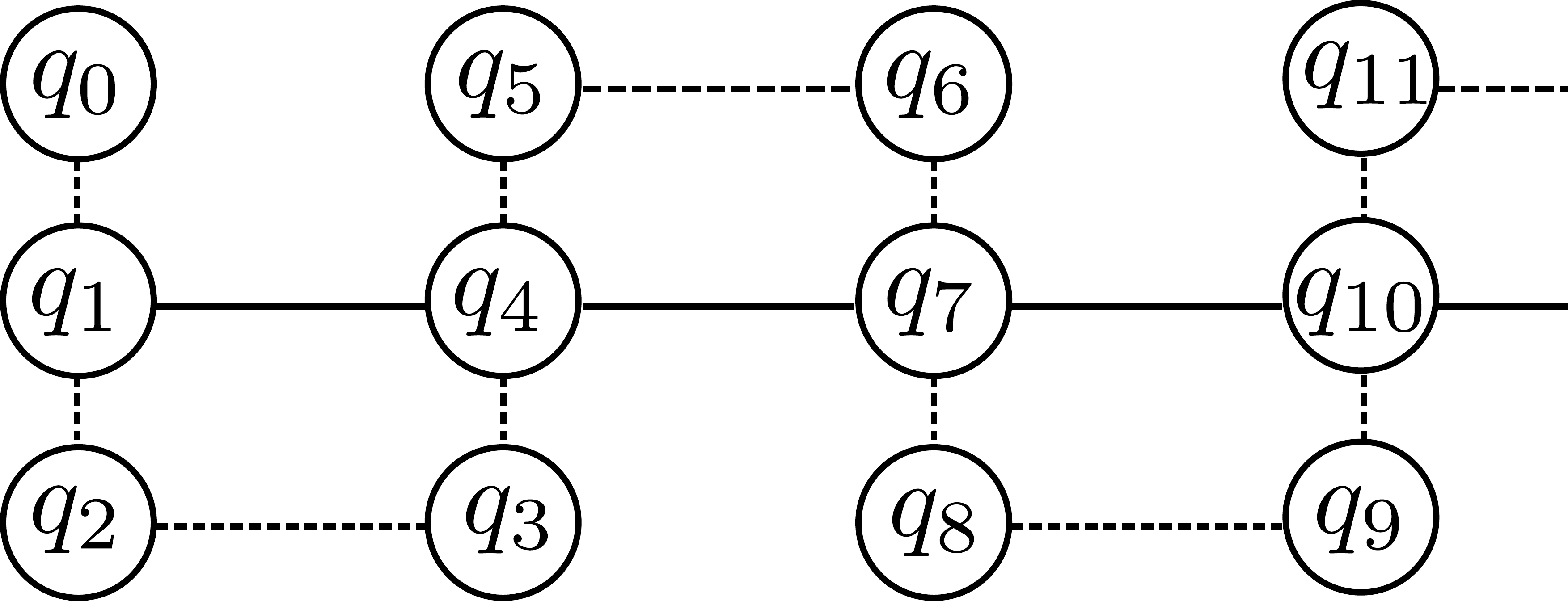}
	\caption{Our algorithm can be implemented using only nearest-neighbor two-qubit gates if the qubit layout of the device contains a three-leg ladder substructure. The solid lines represent couplings between the reduced space registers encoded in the middle sites (stage 1 in Fig.~\ref{fig:circ}). The dashed lines represent the couplings in stage 2 of the algorithm, shown in Fig.~\ref{fig:circ}. }
	\label{fig:layout}
\end{figure}

In an experimental setup, we can apply two-qubit gates only between neighboring qubits. A $3\times n$ square-lattice qubit layout, shown in Fig.~\ref{fig:layout}, makes the middle sites of the consecutive blocks nearest neighbors, allowing for the physical implementation of the circuit.
We present the circuit in Fig.~\ref{fig:circ} as an example for $n=4$. The last two qubits are ``ghost'' sites, which ensure the correct filling fraction. Due to the open boundary conditions of this setup, the properties of the Laughlin state emerge in the bulk of the system and away from the boundaries. We note that with this layout, it is not possible to obtain a lower circuit depth for a deterministic circuit. Since the circuit depth for stage 2 is constant, we focus on stage 1 in the reduced space. With the linear geometry, we need at least an $n-1$ depth to correlate the first and last qubits deterministically.

\begin{figure}
\centering
	\includegraphics[width=8cm]{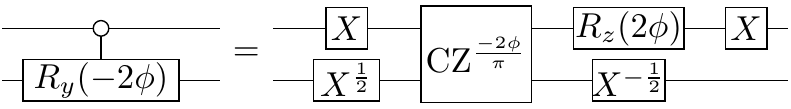}
	\caption{The decomposition of the controlled $R_y$ (up to an overall phase) in terms of native two-qubit gates.}
	\label{fig:elem}
\end{figure}

An efficient decomposition of the controlled $R_y$ gate in term of the native gates of the system is presented in Fig.~\ref{fig:elem}. Our circuit then uses only two types of two-qubit gates, the standard \small{CNOT} and the gate \small{CZ}$^\alpha={\rm diag}(1,1,1,e^{i\pi\alpha})$ for a tunable exponent $\alpha$.

\section{verification of generated Laughlin state}
\label{sec:verification}
We now discuss the measurements that allow us to experimentally verify whether our algorithm successfully prepares the fractional quantum Hall state. We focus on the expectation values of operators acting on the qubits, which can be measured in the existing superconducting quantum devices. As fractional Hall states are topological liquids that do not carry a local order parameter (in real space for the underlying 2D system), such measurements may appear to be challenging. 

However, it has been shown that correlations of some nonlocal string operators would distinguish fractional Hall states from other featureless liquid states. To verify the nature of the state generated, we can perform sets of correlation measurements. The first set consists of one- and two-point correlation functions of the occupation number. While these observables do not serve as an order parameter for topological order, we know their dependence on the parameter $t$. Thus, verification of the expected correlation functions supports the correct preparation of the quantum Hall state.

\begin{figure}[htp]
	\includegraphics[width=8cm]{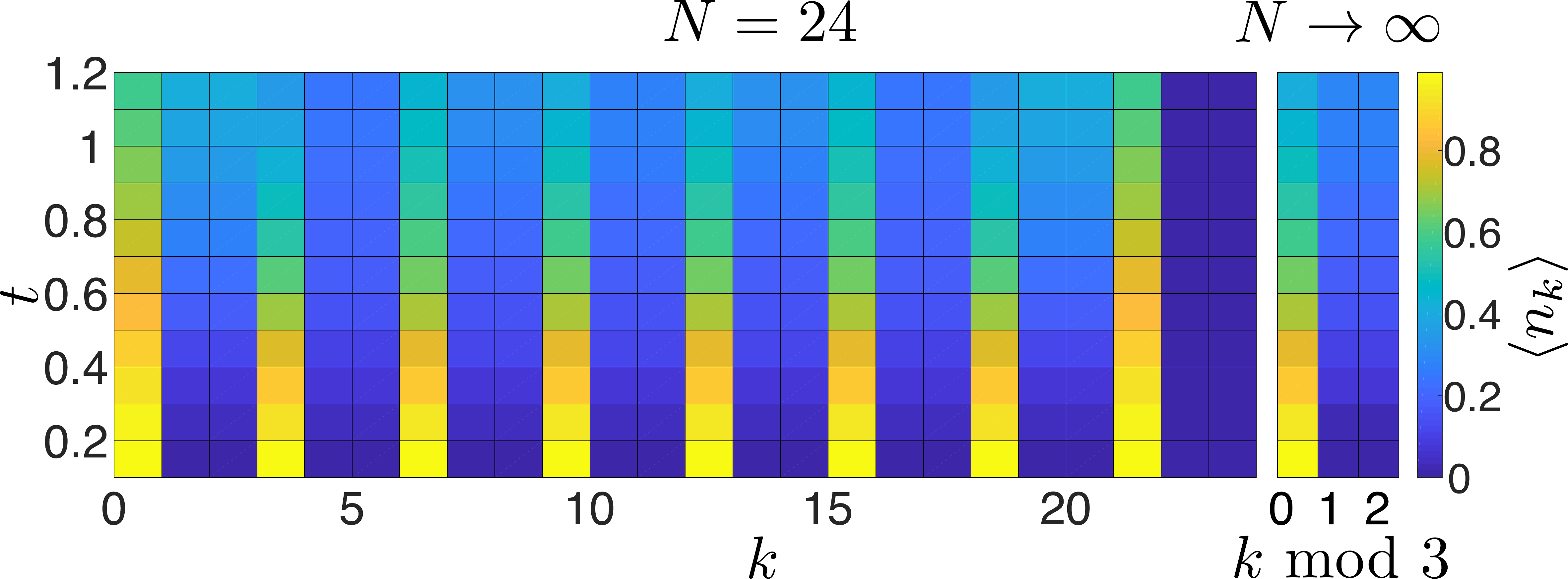}
	\caption{The average density $\langle n_k\rangle$ after the application of our quantum circuit for a system of $N=24$ qubits (the last two qubits are ghost sites that remain zero) and different values of $t$. Away from the boundaries, the results approach the thermodynamic limit expressions given in Eqs. \eqref{eq:n1} and \eqref{eq:n0}. }
	\label{fig:2}
\end{figure}
While we do not have access to the full wave function, as complete tomography is not feasible with a large number of qubits, we can measure $\sigma^z$ for all qubits at the end of the unitary evolution. The local density operator has the simple form $n_j\equiv \frac{1}{ 2}(1-\sigma^z_j)$. The local density pattern and their correlations for the Laughlin state are expected to have the following forms~\cite{nakamura_exactly_2012}

\begin{gather}
\langle n_{3m\pm 1}\rangle=\frac{1}{2}\left(1-1/\sqrt{4t^2+1}\right),\label{eq:n1}\\
\langle n_{3m}\rangle=1/\sqrt{4t^2+1},\label{eq:n0}\\
|\langle n_in_j\rangle-\langle n_i\rangle\langle n_j\rangle|\propto\left(\frac{1-\sqrt{4t^2+1}}{1+\sqrt{4t^2+1}}\right)^{|i-j|/3}\label{eq:nn}
\end{gather}

The results for the density expectation value obtained from our algorithm with a finite number of qubits $N=24$ are shown in Fig. \ref{fig:2}. Already, at this small system size, which can be implemented with 22 qubits, we see that the boundary effects are suppressed in the middle of the chain and the observables are in agreement with the thermodynamic limit expectations.

\begin{figure}[htp]
	\includegraphics[width=8cm]{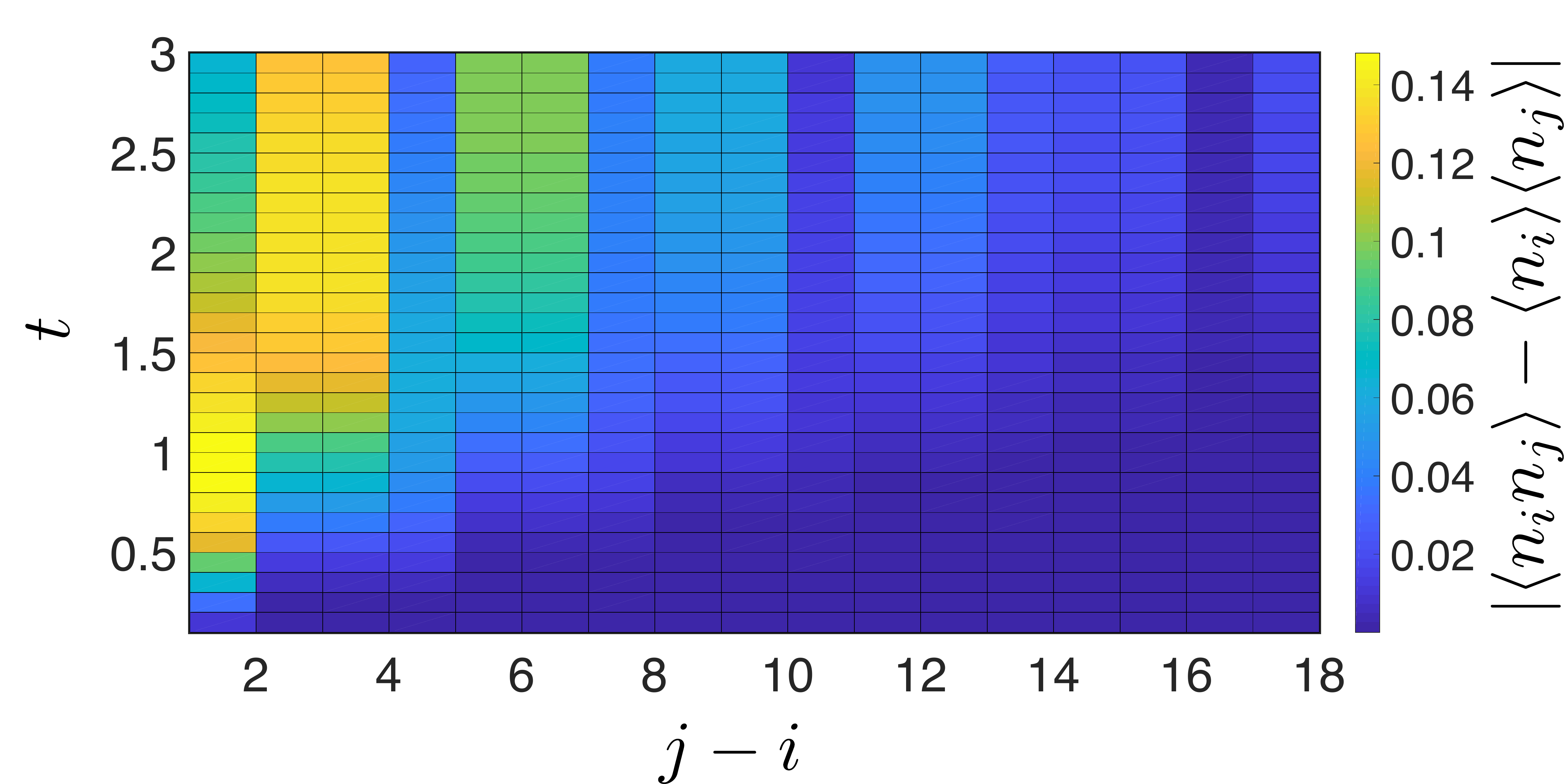}
	\caption{The correlation function $|\langle n_in_j\rangle-\langle n_i\rangle\langle n_j\rangle|$ exhibits fast exponential decay even in a small system with $N=24$ qubits. There are oscillatory finite-size effect, however, which become larger as $t$ increases. In the figure, we use $i=2$ to stay reasonably away from the boundaries.  }
	\label{fig:corr1}
\end{figure}

We can similarly compute the density-density correlation function $|\langle n_in_j\rangle-\langle n_i\rangle\langle n_j\rangle|$. Because we are in the ground state of a gapped Hamiltonian, this correlator decays exponentially with a correlation length that depends on the parameter $t$, as in Eq. \eqref{eq:nn}. In finite systems, it turns out that we can see the dependence of the correlation length on $t$. However, in finite systems there are oscillatory subleading corrections. In Fig. \ref{fig:corr1}, we show the correlator for $i=2$ and several values of $t$ for a system with $N=24$ qubits.

We now turn to a string operator~\cite{denNijs1989}, which can serve as a diagnostic for topological order in this system. Despite the absence of a local order parameter, topological order in a fractional quantum Hall state can be detected by long-range order in the density matrix transformed to a singular gauge~\cite{Girvin1987}.

\begin{figure}[htp]
	\includegraphics[width=8cm]{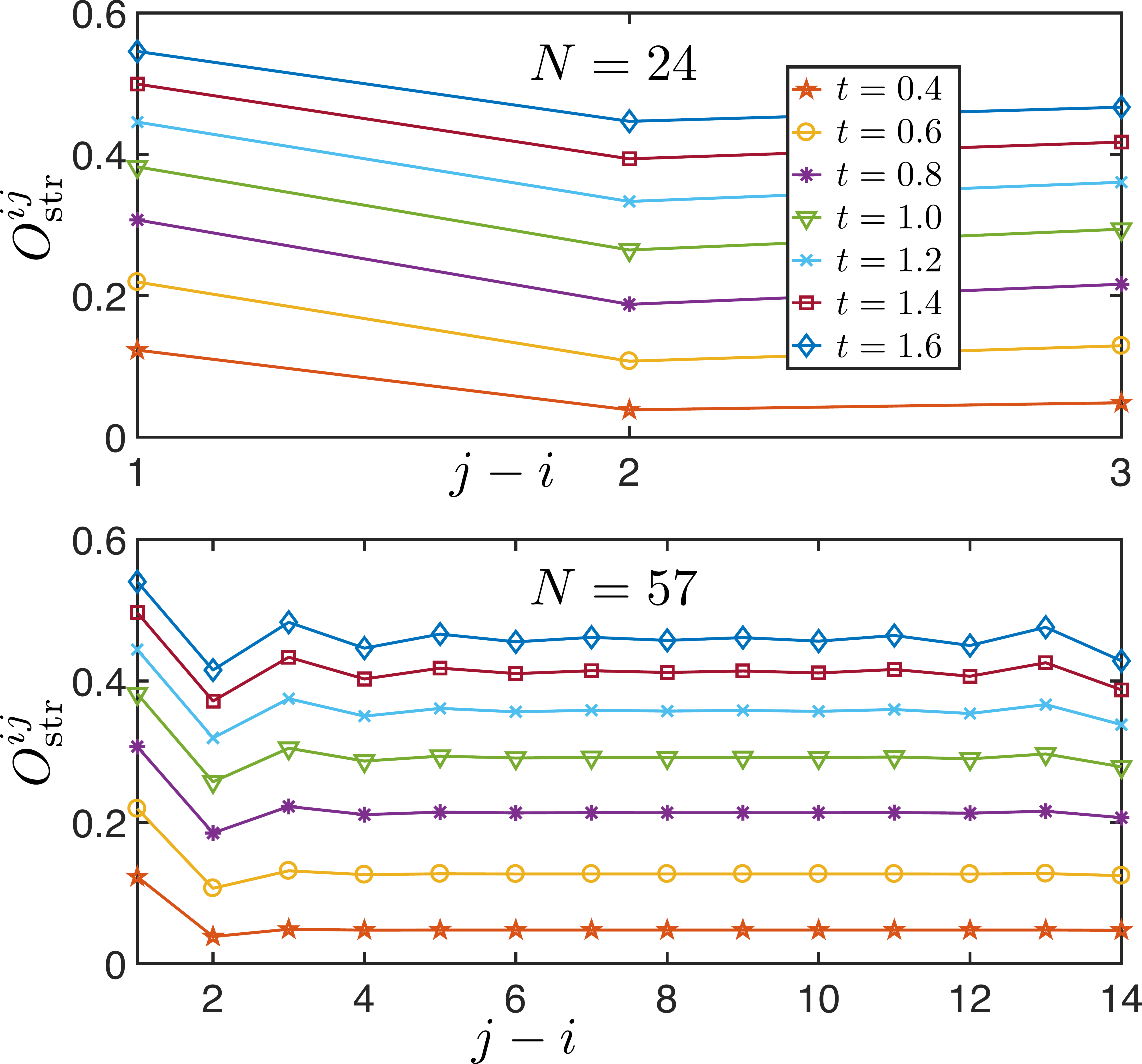}
	\caption{The value of $O^{ij}_{\rm str}$ for $i=1$ as a function of $j-i$ for several values of $t$. The data for small system sizes are strongly suggestive of long-range order. Larger system sizes provide even more compelling evidence.    }
	\label{fig:O_function_of_distance}
\end{figure}

The analog of the singular-gauge density matrix for a spin-1 chain is the string operator  $-\langle S_i^ze^{i\pi \sum_{k=i+1}^{j-1}S_k^z} S_j^z\rangle$ \cite{denNijs1989,Girvin1989}. As in Ref.~\cite{nakamura_exactly_2012}, we identify a spin-1 degree of freedom in our system by noting that the last site of the three-qubit block and the first two sites of the next block can only take three different configurations, $|01,0\rangle\to |S^z=0\rangle$, $|00,1\rangle\to|S^z=1\rangle$, and $|10,0\rangle\to |S^z=-1\rangle$. Therefore, we can write $S_j^z=n_{3j+3}-n_{3j+1}$. We therefore use the following string correlator in terms of the original measured occupation numbers of the qubits:
\begin{equation}
\begin{split}
  O^{ij}_{\rm str}=-\Bigg\langle & \left[\prod_{k=i+1}
  ^{j-1}(-1)^{n_{3k+3}}(-1)^{n_{3k+1}}\right]\\&(n_{3i+3}-n_{3i+1}) (n_{3j+3}-n_{3j+1}) \Bigg\rangle\,.
  \end{split}
\end{equation}
Long-range order in $O_{\rm str}$, i.e., $\lim_{(j-i)\to \infty }O^{ij}_{\rm str}\neq 0$, serves as a diagnostic for the hidden topological order of our state. In Fig.~\ref{fig:O_function_of_distance}, we can clearly observe that for both small and large systems and various values of $t$, the correlation functions do not decay to zero. A qualitative feature of the string operator, namely its finite asymptotic behavior, is an indicator of topological order.
\begin{figure}[htp]
	\includegraphics[width=8cm]{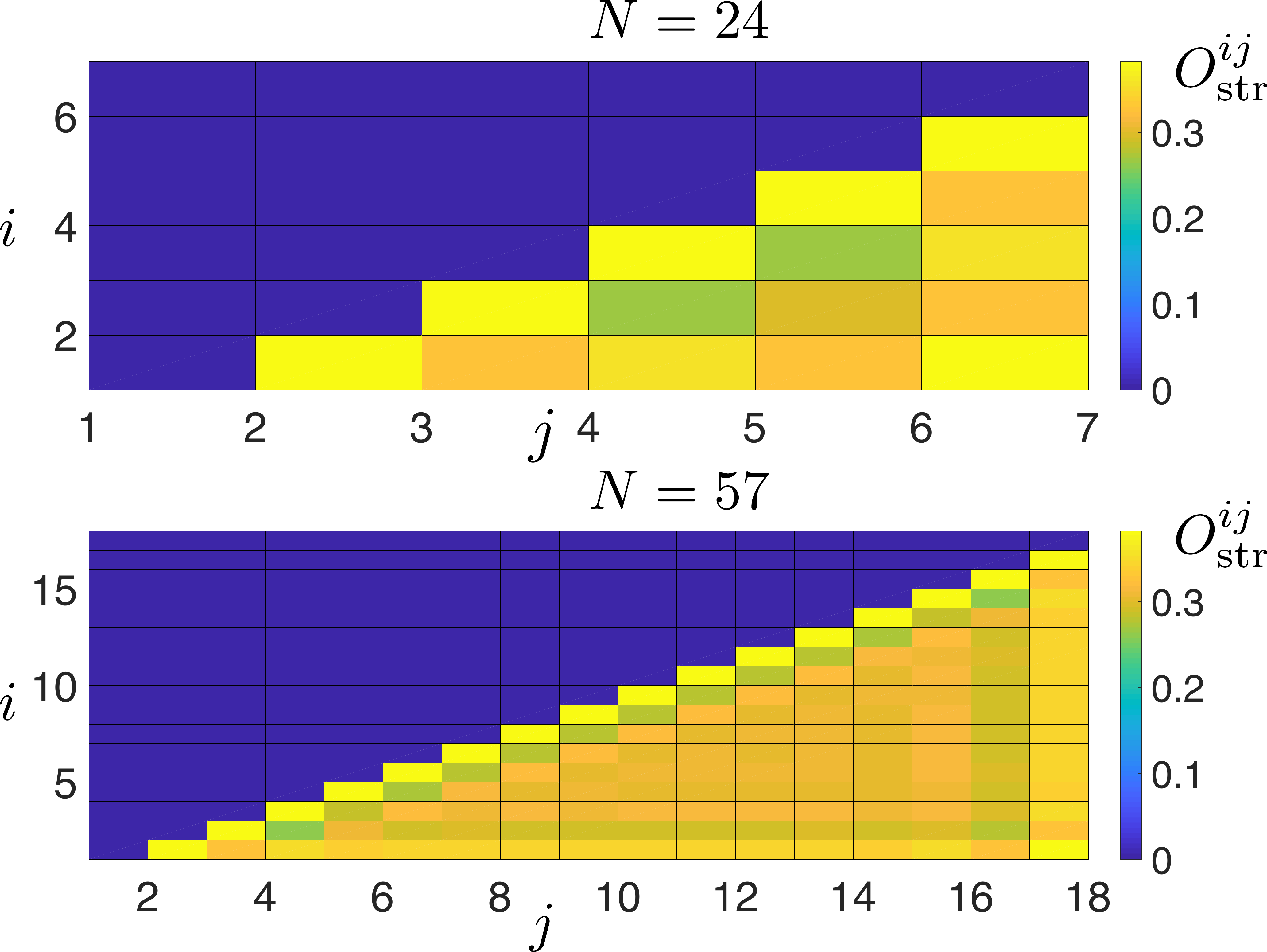}
	\caption{The value of $O^{ij}_{\rm str}$ for $t=1$ and $j>i$ and two system sizes. We have set the correlator to zero for $j\leqslant i$ for easy comparison. Even in a small system, long-range order can be identified. The pattern persists for larger system sizes.}
	\label{fig:O_two_sizes}
\end{figure}

A short comment is in order regarding improving the measurements in experiment. Each measurement results in a bit string and we will need to repeat the experiment many times to estimate the observables $n_j$. In practice, there are errors, such as the $T_1$ error, in the implementation of all the gates. Postselection can correct these errors to a large degree. The squeezing operators do not change the number of ones and zeros, so discarding any bit string for which the number of qubits with $n=1$ is not one third of the total number of qubits improves the precision of the results.

\section{Quasiparticle states }
\label{sec:quasi_particle}
Tbe creation of quasiparticle states is an important challenge, as it can pave the way to studying braiding and the associated topological Berry phases on the quantum device. Using the same recursion relations as before, we can generalize our algorithm to also create quasiparticle states. The Laughlin $\nu=1/3$ state is threefold degenerate. In our discussion so far, we picked the density-wave pattern $100,100,100,\ldots$. Let us represent this pattern by sector $a$. Alternatively, we could use any of the other two thin-torus states $010,010,010,\ldots$ (sector $b$) or $001,001,001,\ldots$ (sector $c$), each of which provides one of the degenerate ground states. Our algorithm works for each of the three sectors above. In each sector, we construct unitary operators that act on four consecutive sites. For the ground state, the blocks that start with 1 have four sites in the 1001 configuration, which is not annihilated by the squeezing operator. The blocks starting with 0, on the other hand, have either the 0010 or 0100 configuration, and get annihilated by squeezing. Therefore, we are able to skip two out of three blocks of four consecutive sites namely, $0010$ and $0100$ in this sector as they are annihilated by the squeezing operators. Our algorithm for sector $a$ therefore implements a unitary operator 
\begin{equation}
    {\cal U}^a=U_{N/3}^a U_{N/3-1}^a\cdots U_1^aU_0^a,
\end{equation}
where $U^a_k$ is the same as $U_k$ defined in Eq. (\ref{eq:uk}).
Similarly, we can define
\begin{equation}\label{eq:ukbc}
    U_k^b=e^{\phi_k(S_{3k+1}-S^\dagger_{3k+1})},\quad U_k^c=e^{\phi_k(S_{3k+2}-S^\dagger_{3k+2})}.
\end{equation}
A more general unitary operator
\begin{equation}
    {\cal U}=U_{N/3}^cU_{N/3}^bU_{N/3}^a \cdots U_0^cU_0^bU_0^a
\end{equation}
can be constructed, which can prepare the ground state in all three sectors. Quasiparticles can be constructed by acting with $\cal U$ on density-wave states that have domain walls between patterns corresponding to the different sectors. The above unitary, which acts on all qubits, correctly implements the transformation from thin- to thick-torus limits, and works also in the presence of domain walls.

\section{Conclusions}
\label{sec:conclusion}
In summary, we design an efficient quantum algorithm for creating a $\nu=1/3$ Laughlin-type wave function in a system of qubits arranged in a three-leg ladder. Only single-qubit gates and two-qubit gates acting on nearest neighbors in the ladder geometry are used in our algorithm.  The circuit depth is $N/3 + 3$, which scales linearly with system size.

Our algorithm takes advantage of the matrix-product nature of the Laughlin-type wave functions in the second quantization picture using Landau orbitals. While the assignment of one creation or annihilation operator to each orbital may hide some characteristic properties of the underlying 2D fractional quantum Hall state, the topological order of the fractional Hall effect can still be diagnosed by measuring some string operators. 

The physical creation of the Laughlin wave function in a ladder geometry paves the way for several important studies. These include quantum simulation of nonequilibrium dynamics of quantum Hall states, as well as braiding quasiparticles by implementing a cyclic adiabatic process on the spin chain with domain walls. Due to the short circuit depth and the easy procedure to verify the generated state, our algorithm is well positioned to be implemented on near-term quantum computers. 

\section{acknowledgements} We would like to thank Maissam Barkeshli, Mohammad Hafezi, Tom O'brien, Emil Prodan and Ryan Babbush for helpful discussions. This work was supported by Google AI lab. AR acknowledges support from National Science Foundation Awards DMR-1945395 and DMR-2038028. AR is grateful to the Kavli Institute for Theoretical Physics for hospitality
during parts of this project, acknowledging support by the National Science
Foundation under Grant No. NSF PHY-1748958. PG acknowledges support from National Science Foundation Awards DMR-1824265 and DMR-2037996. 

%
\bibliographystyle{apsrev4-1_with_title}
\bibliography{fhm}


\end{document}